\documentclass[useAMS,usegraphicx,usenatbib]{mn2e}  
\usepackage{epsfig}
\usepackage{tikz}

\def\micron{\,$\mu$m\,}
\def\deg{$^{\circ}$}
\def\aap{A\&A}
\def\apj{ApJ}

\def\apjs{ApJS}
\def\mnras{MNRAS}

\def\aj{AJ}
\def\pasp{PASP}

\title[Star formation with Galactocentric radius]{The prevalence of star formation as a function of Galactocentric radius}
\author[S. E. Ragan et al.]{S. E. Ragan$^{1}\thanks{email: S.Ragan@leeds.ac.uk}$, T. J. T. Moore$^{2}$, D. J. Eden$^{2}$, M. G. Hoare$^{1}$, \and D. Elia$^{3}$, S. Molinari$^{3}$\\
\\ $^{1}$ School of Physics and Astronomy, University of Leeds, Leeds, LS2 9JT, UK \\
$^{2}$ Astrophysics Research Institute, Liverpool John Moores University, IC2, Liverpool Science Park, 146 Brownlow Hill, Liverpool, L3 5RF, UK \\
$^{3}$ INAF-Istituto di Astrofisica e Planetologia Spaziale, Via Fosso del Cavaliere 100, I-00133 Roma, Italy 
}

\begin{document}
\maketitle

\label{firstpage}

\begin{abstract}
We present large-scale trends in the distribution of star-forming objects revealed by the Hi-GAL survey. As a simple metric probing the prevalence of star formation in Hi-GAL sources, we define the fraction of the total number of Hi-GAL sources with a 70\micron counterpart as the ``star-forming fraction'' or SFF. The mean SFF in the inner galactic disc (3.1\,kpc$\,< R_{\rm GC}\,<$ 8.6\,kpc) is 25\%. Despite an apparent pile-up of source numbers at radii associated with spiral arms, the SFF shows no significant deviations at these radii, indicating that the arms do not affect the star-forming productivity of dense clumps either via physical triggering processes or through the statistical effects of larger source samples associated with the arms. Within this range of Galactocentric radii, we find that the SFF declines with $R_{\rm GC}$ at a rate of $-$0.026$\pm$0.002 per kiloparsec, despite the dense gas mass fraction having been observed to be constant in the inner Galaxy. This suggests that the SFF may be weakly dependent on one or more large-scale physical properties of the Galaxy, such as metallicity, radiation field, pressure or shear, such that the dense sub-structures of molecular clouds acquire some internal properties inherited from their environment.  
\end{abstract}

\begin{keywords}
galaxies: ISM -- ISM: clouds --  stars: formation
\end{keywords}

\section{Introduction}
Molecular gas is a dominant component in the interstellar medium (ISM) and the principal location of star formation. Clouds of molecular gas take on a hierarchical structure throughout the Milky Way, where the densest clumps of gas account for roughly 5-10\% of the total mass in a typical cloud \citep{Battisti2014,Ragan2014}. With the exception of the central molecular zone (CMZ), this appears to be a universal property of molecular clouds on average, regardless of a cloud's proximity to a spiral arm where, globally, molecular gas is concentrated \citep{Eden2012,Eden2013}.  

The conditions of gas in molecular clouds in the Milky Way is often characterised by CO emission, and large sections of the Galactic plane have now been surveyed in a number of CO transitions and isotopologues. For example, \citet{Roman-Duval2010} use the Galactic Ring Survey (GRS) $^{13}$CO (1-0) data and find a steep decline in the Galactic surface mass density of molecular clouds with Galactocentric radius ($R_\mathrm{GC}$), which extends to the outer Galaxy until a truncation point of the molecular disc at $R_\mathrm{GC}$ = 13.5\,kpc \citep{Heyer1998}. 
The excitation temperature of CO declines with $R_\mathrm{GC}$, which may link to interplay between the slow decline of the cooling rate (due to lower metallicity) and more rapid decline of the heating rate (attributed to a decrease in star formation rate [SFR]) with $R_\mathrm{GC}$ \citep{Roman-Duval2010}.

Thermal emission from interstellar dust grains which follow the gas distribution provide a secondary tracer of ISM structure and properties. \citet{Sodroski1997} derive a dust temperature gradient with $R_\mathrm{GC}$, owing in part to the metallicity gradient in the Galactic disc \citep{Lepine2011b} and to variations in the strength of the interstellar radiation field with $R_\mathrm{GC}$. The links between these trends and star formation, however, remain tenuous.

Data from the {\em Herschel} Hi-GAL survey \citep{Molinari2010b} provides a new high-resolution perspective on the distribution of dust which is necessary to distinguish between active and quiescent molecular clouds throughout the Galaxy and to quantify their star formation activity in detail. In this paper, we examine global trends in the properties of Hi-GAL sources with Galactocentric radius.  

\begin{figure}
\includegraphics[width=0.5\textwidth]{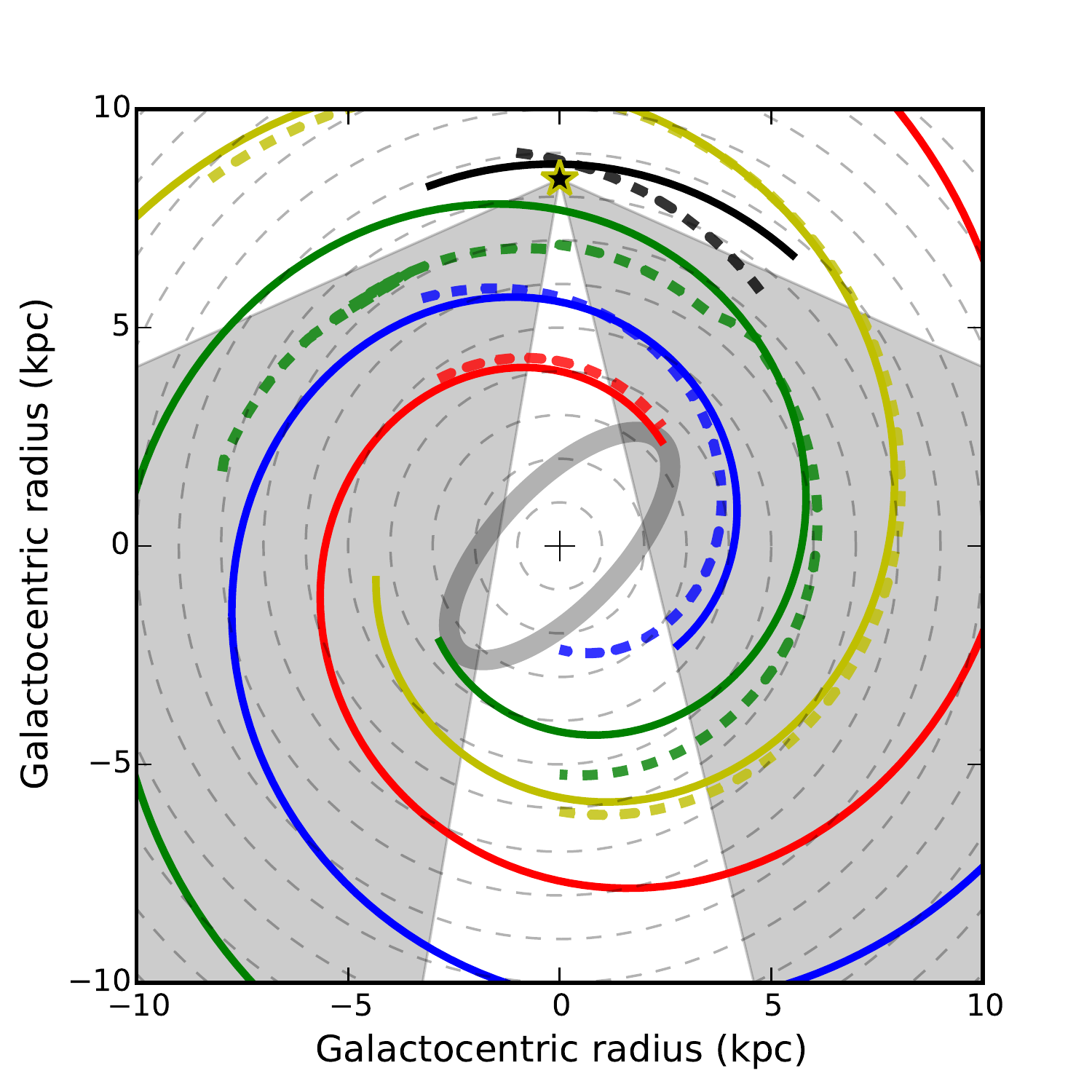}
\caption{\label{f:hou_spiral} A bird's eye schematic view of the Milky Way. The solid coloured spiral features correspond to the \citet{HouHan2014} four-arm spiral arm model based on HII regions, where red is the Norma arm, yellow is the Perseus arm, green is the Sagittarius-Carina arm, blue is the Scutum-Centaurus arm, and black is the local arm.  The grey ellipse represents the area influenced by the Galactic bar. The dashed lines show the \citet{Reid2016} loci of the spiral arms, colour-coded in kind as above. The star indicates the position of the Sun. The grey shaded regions show the area bound by the longitude limits of the current catalogue. }
\end{figure}

\section{Data}

The {\it Herschel} key program Hi-GAL \citep{Molinari2010b,Molinari2010a} surveyed the plane of the Milky Way in five photometric bands available with the PACS \citep[70 and 160\micron;][]{A&ASpecialIssue-PACS} and SPIRE \citep[250, 350 and 500\micron;][]{A&ASpecialIssue-SPIRE} instruments.  These wavelengths cover the peak of the spectral energy distribution of thermal emission from dust grains in the temperature range 8\,K $< T_{\rm dust} <$ 50\,K. Compact sources at these wavelengths represent the regions in the Galaxy which have the cold, dense conditions necessary for star formation. 

We use the Hi-GAL compact source catalogue \citep{Molinari2016b}, which covers the inner Galaxy longitudes of 14\deg $<$ $l$ $<$ 67\deg and 293\deg $<$ $l$ $<$ 350\deg. A schematic region of the Galaxy covered is shown in Figure~\ref{f:hou_spiral}. The spiral arms show the analytic four-arm model from \citet[][hereafter HH14]{HouHan2014} fit to HII regions (assuming $R_0 = 8.5$\,kpc and $\Theta_0 = 220$\,km\,s$^{-1}$) and the model from \citet[][hereafter R16]{Reid2016}. We will discuss the differences between spiral arm models in Section\,4.2. Catalogue sources from each band were matched according to the method described in \citet{Giannini2012}, which uses spatial associations starting from 500\micron and moving toward shorter wavelengths (a band-merged source catalogue will be published separately; Elia et al. in preparation). 
For the following, we require sources to be detected in at least three adjacent bands -- either 160, 250 and 350\micron or 250, 350 and 500\micron -- and to have kinematic distance estimates derived using the rotation-curve-based methods described in \citet{Russeil2011}, assuming the distance between the Sun and Galactic centre to be 8.4\,kpc. This results in 57077 sources.

The presence of 70\micron emission is a reliable indicator of embedded star formation activity \citep[e.g.][]{Ragan2012b,Traficante2015}. \citet{Molinari2016b} indicate that the 70\,\micron Hi-GAL catalogue is complete to $\sim$95\% above 0.5\,Jy, which we adopt as the threshold for a source to qualify as 70\micron-bright. Due to differences in angular resolution across the different Hi-GAL bands, sources detected at longer wavelengths may be associated with more than one 70\micron source. This affects $\sim$5\% of the catalogue entries, and we take the sum of the 70\micron flux from all sources associated with the entry. 
In the following we explore the prevalence of Hi-GAL sources with 70\micron counterparts compared to the distribution of all sources. Hereafter we will refer to the fraction of Hi-GAL sources with a 70\micron counterpart as the ``star-forming fraction'' or SFF. 

\section{Galactic scale trends}

\begin{figure}
\includegraphics[width=0.5\textwidth]{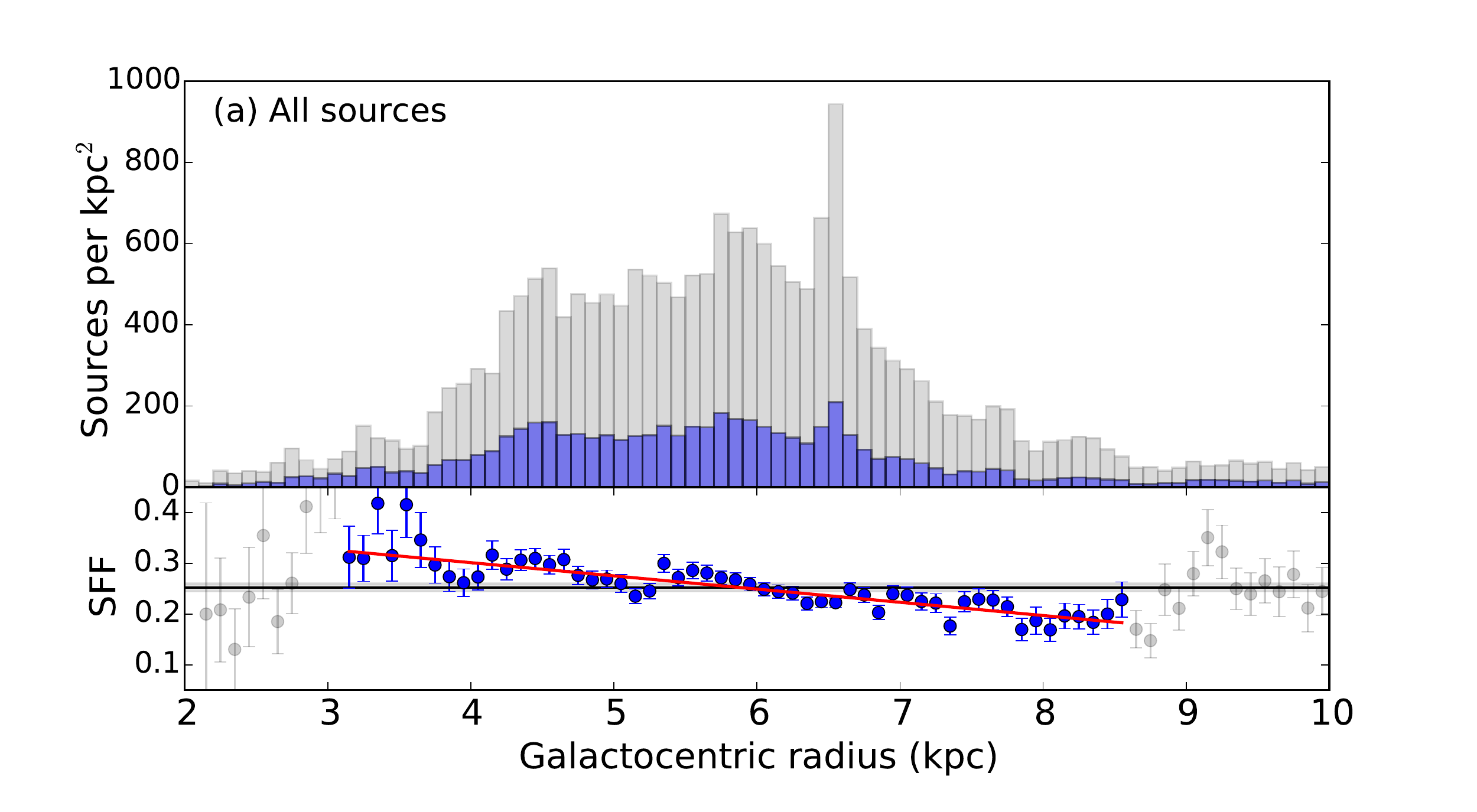} \\
\vspace{-0.1in}
\includegraphics[width=0.5\textwidth]{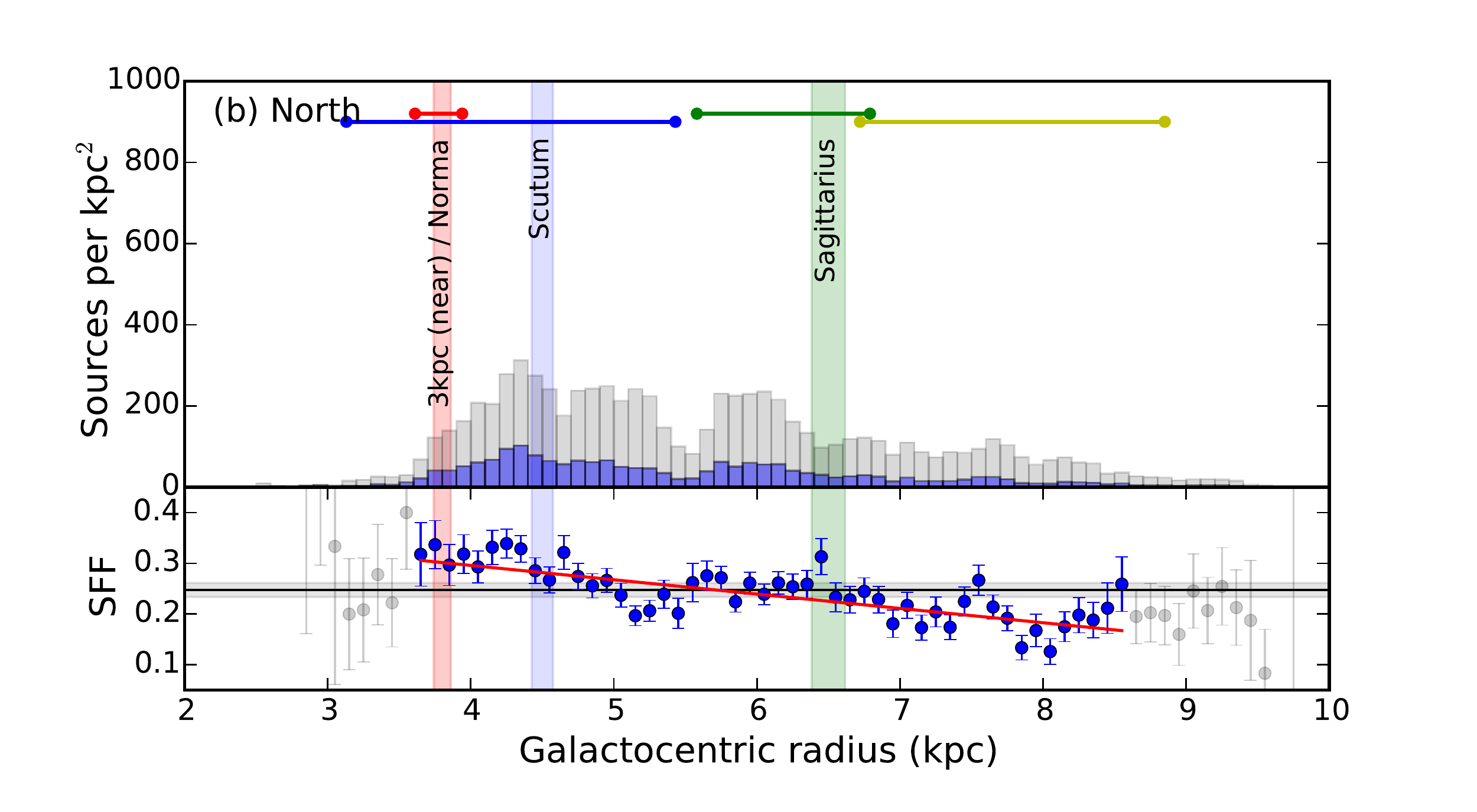} \\\vspace{-0.15in}
\includegraphics[width=0.5\textwidth]{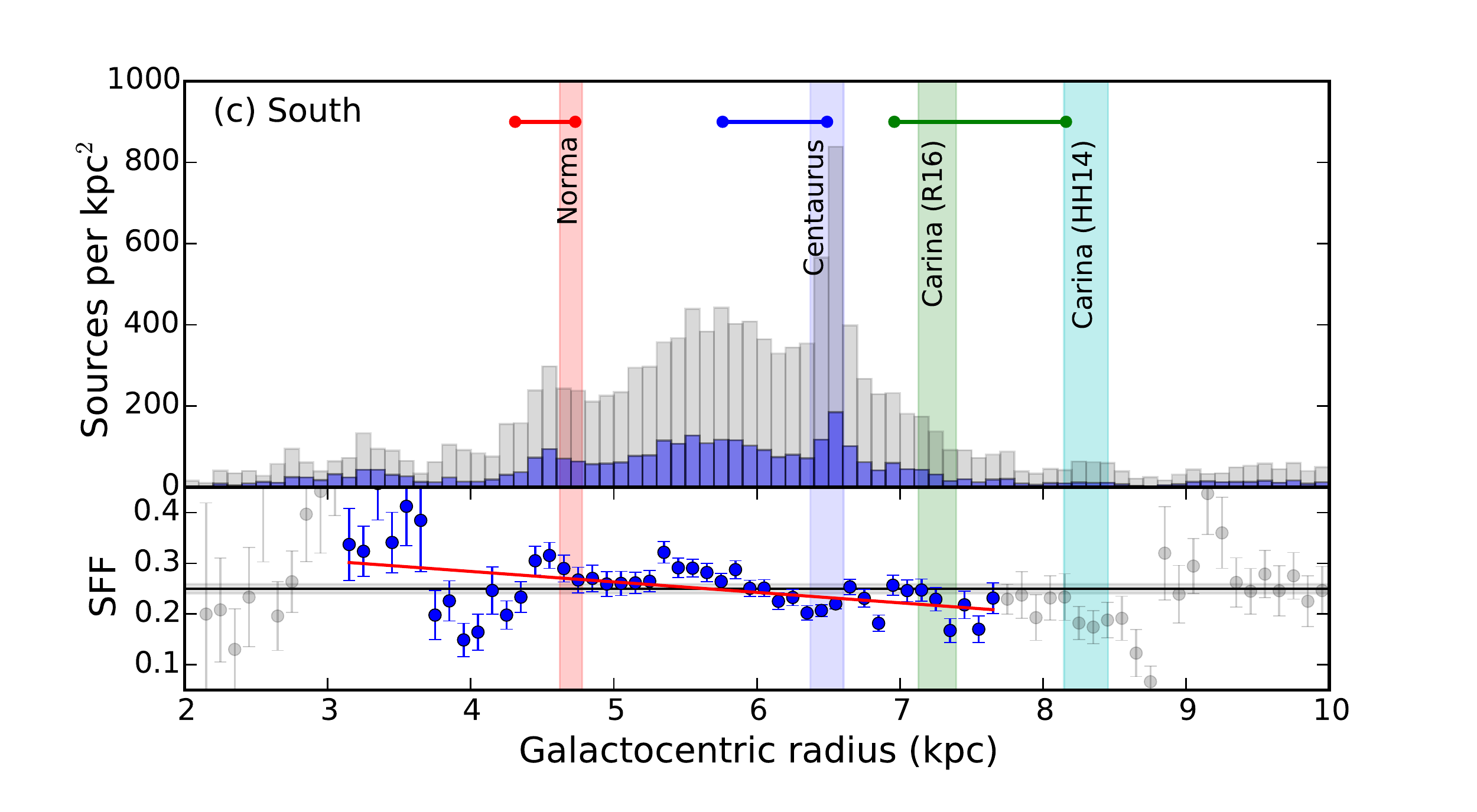} 
\caption{{\bf (a)} {\it Top panel:} Histogram of total number of Hi-GAL sources per unit area as a function of the Galactocentric radius (in kiloparsecs) overlaid (blue) with the subset of these sources with  70\,$\mu$m counterparts ($S_{\nu} >$ 0.5\,Jy). {\it Bottom panel: } The fraction of the total sources with 70\,$\mu$m counterparts. The error bars reflect the Poisson statistical errors. Blue points are bins which have $>$100 total sources and were used in the fits; grey points denote bins with $<$100 sources and were excluded. The solid horizontal line at SFF = 0.25 is the mean fraction of the sample, and the standard error of the mean is shown in the grey shaded region. The weighted linear least squares fit is shown over the range of Galactocentric radius bins with 100 or more sources in the red solid line.
{\bf (b)} Same as (a), but for the northern subsample. The coloured horizontal lines show the range of $R_{\rm GC}$ spanned by each spiral arm (red: 3kpc (near) / Norma; blue: Scutum; green: Sagittarius-Carina; yellow: Perseus) in our longitude range. The vertical blue and green shaded bars correspond to the approximate locations of the tangent points to the Scutum and Sagittarius arms; the red shaded bar marks the end of the near side of the 3\,kpc arm and start of the Norma arm. These reference points are in rough agreement between the two spiral arm models considered. The widths of these regions reflect the $R_\mathrm{GC}$-dependent arm width calculated in \citet{Reid2014a}. 
{\bf (c)} Same as (a), but for the southern subsample. The coloured horizontal lines show the range spanned by each arm (see above for [b]), and the vertical shaded regions show the location (and arm width) of the tangents to each arm (red: Norma; blue: Centaurus; green: Carina \citep{Reid2016}; cyan: Carina \citep{HouHan2014}). \label{f:Nsources_area_Rgc}}
\end{figure}

We show the overall distribution of Hi-GAL sources in the grey histogram in the upper panel of Figure~\ref{f:Nsources_area_Rgc}a, and again in the upper panels of Figures~\ref{f:Nsources_area_Rgc}b and \ref{f:Nsources_area_Rgc}c for the North (14\deg $< l <$ 67\deg) and South (293\deg $< l <$ 350\deg), respectively. The counts in the 0.1\,kpc-wide bins are normalised by the area of the Galactocentric annuli at the given radius, also accounting for the longitude limits of the catalogue. 

We show the subset of sources with 70\micron counterparts with the blue histograms in each upper panel of Figure~\ref{f:Nsources_area_Rgc}. We find that within the area covered by the current catalogue (i.e. the inner Galaxy excluding the CMZ), the overall mean SFF is 0.25, very similar to the fraction of Bolocam Galactic Plane Survey (BGPS) sources with associated young stellar objects (0.29) reported by \citet{Eden2015}. 

The lower panels of each plot in Figure~\ref{f:Nsources_area_Rgc} show the fraction of the total number of sources that are bright at 70\micron, and the errorbars represent the propagated Poisson errors for the counts of each population per bin. The blue points indicate bins which have more than 100 total sources, which were used in the following analysis.

The distributions shown in Figure~\ref{f:Nsources_area_Rgc} exhibit no significant peaked deviations from the mean SFF, however a gradual declining trend in SFF with Galactocentric radius is evident. The Spearman rank correlation coefficient for the total distribution (Figure~\ref{f:Nsources_area_Rgc}a) is $\rho_S$ = $-$0.91, thus we can conclude that a highly significant monotonic trend exists.

We perform a weighted least squares linear fit to the relation between SFF and $R_{\rm GC}$ using Poisson error weighting. We restrict the fit to $R_\mathrm{GC}$ bins with at least 100 counts and find the following expression as the best fit:

\begin{equation}
{\rm SFF} = 
(-0.026\pm 0.002)~R_\mathrm{GC} + (0.406\pm 0.003)\  [{\rm kpc}^{-1}]
\end{equation}

\noindent The overall negative gradient is robust in the northern and southern subsamples, with slopes of $-$0.030$\pm$0.004 per kpc in the north (fit where $N > 100$ per 0.1\,kpc-wide bin, 3.6 $<$ $R_\mathrm{GC}$ $<$ 8.6\,kpc) and $-$0.021$\pm$0.006 per kpc in the south (fit between 3.1 $<$ $R_\mathrm{GC}$ $<$ 7.7\,kpc).

By separating the Hi-GAL sources into the northern and southern Galactic subsamples, some of the spiral arm features in the $R_{\rm GC}$ distributions are recovered. In Figures~\ref{f:Nsources_area_Rgc}b and \ref{f:Nsources_area_Rgc}c, we show the average positions of the spiral arms within the longitude limits according to the most recent spiral arm models from the BeSSeL project \citep{Reid2016}, reflecting the $R_{\rm GC}$-dependent arm widths \citep{Reid2014a}. 
In the top panel of Figure~\ref{f:Nsources_area_Rgc}c, there is a clear peak in the total source distribution at $R_{\rm GC} \sim$6.6\,kpc which likely corresponds to the Centaurus arm, and the  peak at 4.5\,kpc corresponds to the Norma arm in our longitude range. Despite these arm features appearing in the overall distribution (top panels), the SFF (bottom panels) do not show any significant peaks ($> 3\sigma$ above mean) at these radii. 

Our linear weighted least squares fit to the full sample (Figure~\ref{f:Nsources_area_Rgc}a) results in a slope of $-$0.026$\pm$0.002 per kpc within the 3.1 $<$ $R_\mathrm{GC}$ $<$ 8.6\,kpc range, which is of 10-$\sigma$ significance. However, this fit is weighted with the Poisson error for each bin, which may not adequately account for the dominant uncertainty in the distance estimate. So we next test the significance considering the large distance uncertainties. For simplicity, we first assume a uniform fractional error for each distance estimate, i.e. the distance to each source is certain within $\pm$ 10\%, 20\%, 30\%, 40\% or 50\%. Next, for each individual source in the catalogue, we simulate a new $R_\mathrm{GC}$ by randomly sampling within  a normal distribution of width the size of the assumed errorbar. With the set of new distances for all catalogue entries, we perform a weighted least squares linear fit. We repeat this exercise 10000 times and record the new slope for each simulation.  

\begin{figure}
\includegraphics[width=0.5\textwidth]{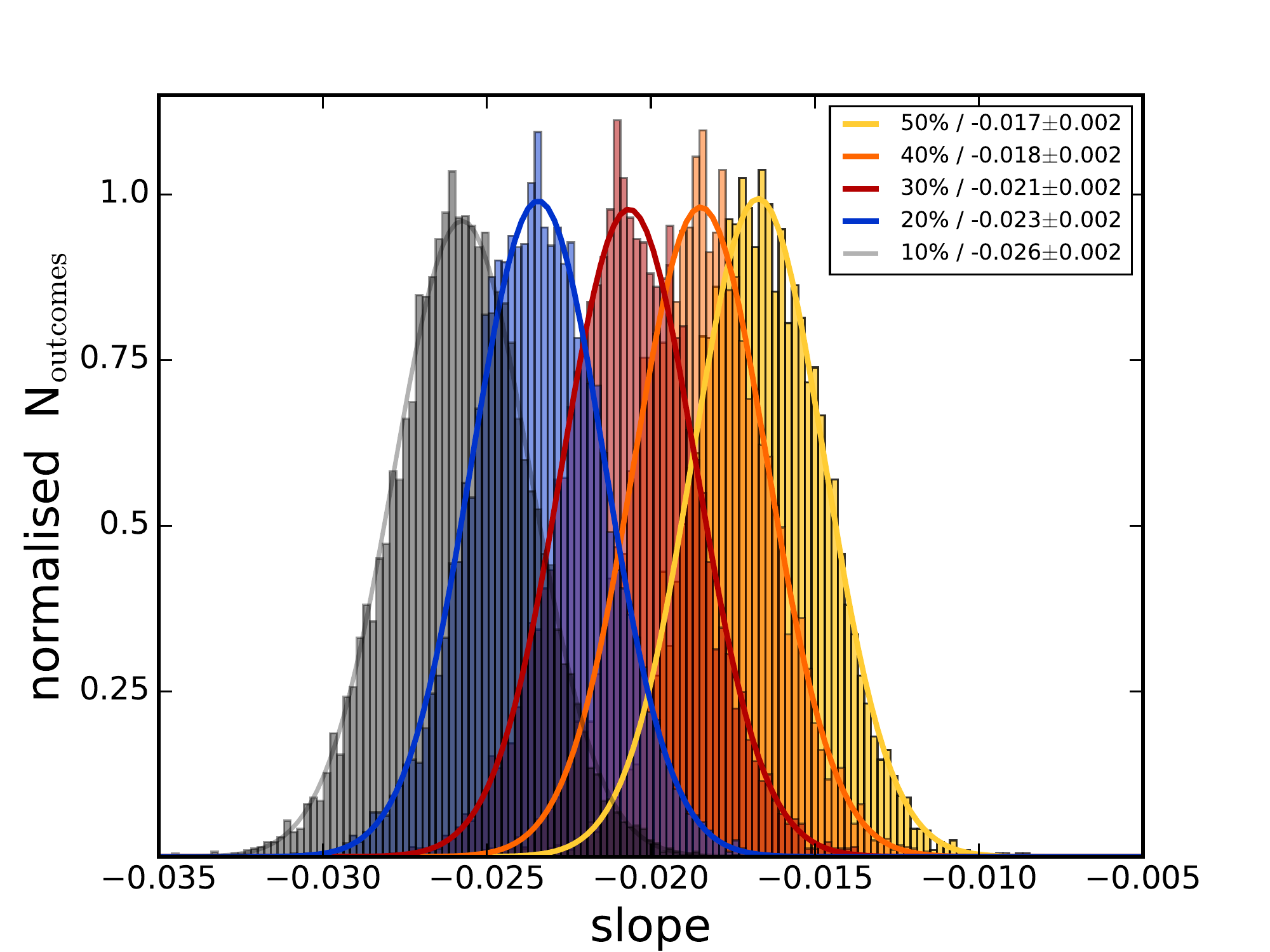}
\caption{Distribution of slopes when randomly sampling (N=10000 times) within the distance estimation error. Different fractional errors are shown in the different colours: 50\% (yellow), 40\% (orange), 30\% (red), 20\% (blue) and 10\% (grey). The key shows the mean and standard deviation of each distribution. \label{f:simulated_slopes}}
\end{figure}

The mean and dispersion of the simulated slope distributions of the SFF versus $R_\mathrm{GC}$ relation assuming a 50\% / 40\% / 30\% / 20\% / 10\% error in heliocentric distance is still a significant result in all cases. We summarise the distributions of the slopes in Figure~\ref{f:simulated_slopes}, where the slope and 1-$\sigma$ error are given for each fractional error assumption.

\section{Discussion}

\subsection{What does the star-forming fraction (SFF) mean?}
Hi-GAL sources detected at the four longest {\em Herschel} wavebands -- 160, 250, 350 and 500\,\micron -- are equivalent to the submillimetre-continuum sources detected in surveys such as ATLASGAL \citep{ATLASGAL} or the JCMT Plane Survey \citep[JPS;][]{Moore2015}. For the most part, such objects have virial ratios clustered around the critical value \citep{Urquhart2014c} and therefore at least half are potentially star-forming clumps. As explained above, we take it that those that have 70\,\micron emission are already actively star-forming. The SFF may therefore be considered as the fraction of dense clumps with embedded YSOs. If all the IR-dark clumps detected by Hi-GAL were to evolve into IR-bright sources, the SFF would give the relative timescales of the pre-stellar and protostellar stages, but this is not necessarily the case. Those not currently forming stars may either go on to form stars in the future or may dissipate without doing so.

The SFF must be somewhat related to the evolutionary state of clumps, as traced by e.g. the ratio of their infrared luminosity ($L_{\rm IR}$) to mass \citep[cf.][]{Molinari2008, Urquhart2013a}, and has some dependence on the average clump mass, since the highest-mass clumps have an undetectably short infrared-dark lifetime \citep{Motte_cygX,Urquhart2014c}.
The mean SFF is therefore set by the relative timescale of the IR-bright protostellar stage to that of the pre-stellar stage, multiplied by the average fraction of productive dense cores. The measured mean SFF value of 0.25 happens to be consistent with equal timescales and 50\% of clumps being eventually star-forming \citep[see also][]{Moore2015}. Relative variations in SFF indicate changes (in both, but presumably mainly the latter) and/or variations in the time gradient of the SFR on timescales similar to the clump lifetime ($\sim 10^5$ years; if SFR is increasing, SFF will be low, since there will be more bound starless clumps, and vice versa). 

The SFF is therefore a quantity related to the current star-formation efficiency (SFE) within dense, potentially star-forming clumps, being the fraction of dense clumps that are forming stars within the timescale set by submillimetre and far-IR detection. 
The SFF does not, however, tell us the actual conversion efficiency of clump mass into stellar mass, and is thus not a star-forming {\it efficiency}, strictly speaking. 
A change in SFF with location may indicate a spatial variation in environmental factors influencing the probability that a clump will form stars. Such factors may include the  availability of dense molecular gas \citep{Roman-Duval2016}, turbulent pressure \citep{Wolfire2003}, local magnetic-field strength \citep{HeilesTroland2005}, or the presence of a triggering agent such as a wind- or radiation-driven bubble \citep{Bertoldi1989,Bisbas2009}.

\subsection{The SFF associated with spiral arms}

Our knowledge of the spiral structure of the Galaxy comes from high-resolution surveys of the Milky Way plane, which have informed various efforts to model Galactic structure \citep[e.g.][]{Dame2001,Roman-Duval2010,Vallee2014c,HouHan2014,Reid2014a,Reid2016}. The Milky Way has either two or four arms \citep[depending on the choice of tracer;][]{Robitaille2012} and exhibits spatial offsets between tracers of cold molecular gas and those of active star formation \citep[e.g.][]{Vallee2014a}. While these studies have shown that the spiral arms are undoubtedly where material is concentrated in the Galaxy, studies of SFE metrics across the Galactic plane have found no compelling evidence of variation associated with the spiral arms \citep{Moore2012,Eden2012,Eden2013,Eden2015}, though small number statistics were a limitation to these studies. 

Nevertheless, we find similar signatures in our results. Figure~\ref{f:Nsources_area_Rgc} shows that in terms of total number of sources (top histogram panels), there are enhancements at the spiral arm radii. In the north, the Scutum tangent at $R_{\rm GC} \sim 4.5$\,kpc, the Sagittarius arm between $5.5 < R_{\rm GC} < 6.5$\,kpc show clear peaks in total distribution. The southern Norma ($R_{\rm GC} \sim 4.7$\,kpc) and Centaurus ($R_{\rm GC} \sim$ 6.5\,kpc) tangents are also peaks in total source surface density. One of the largest discrepancies between competing spiral arm models is the path of the Carina arm. If it lies between 7\,kpc $< R_{\rm GC} <$ 8\,kpc as R16 suggest, is not a peak in overall source surface density or SFF. The HH14 model puts the Carina arm about 1-2\,kpc further from the Galactic centre in \ $R_{\rm GC}$, in which case the current catalogue misses the tangent longitude and  the Carina arm is too poorly-sampled for our consideration in this paper.

 Turning to the lower panels in Figure~\ref{f:Nsources_area_Rgc} (SFF versus $R_{\rm GC}$), we see that the SFF at these radii do not exhibit compelling peaks (i.e. $>3\sigma$ deviation from the mean SFF), with the possible exception of one bin near the northern Sagittarius arm tangent ($R_{\rm GC} \sim$ 6.5\,kpc) where the SFF is $\sim$0.31 ($\sim$2$\sigma$), however since the adjacent bins lack any elevation in SFF, this peak should be taken cautiously. Otherwise and interestingly, if anything, the SFF exhibits weak depressions in SFF at the Perseus, (southern) Centaurus and Carina arm radii. 

That the Sagittarius arm (at $R_\mathrm{GC} \sim$ 6 -- 6.5\,kpc in the North) may be unremarkable in SFF versus $R_{\rm GC}$ is of particular interest. This arm is prominent in CO\,(3-2) and therefore has abundant molecular gas content at these longitudes \citep{Rigby2016}. It is also a strong feature in the RMS source distribution \citep{Urquhart2014a}, gas temperature \citep{Roman-Duval2010}, and the ratio of infrared luminosity to clump mass \citep[$L_{\rm IR}/M_{\rm clump}$][]{Moore2012,Eden2015} suggestive of enhancement in the SFE, albeit on the kiloparsec scales probed by earlier surveys. This, however, may be a consequence of local variations (e.g. a few high-luminosity sources) which are not captured by the SFF metric, which is based strictly on source count surface densities.

The lack of significant change in SFF across the spiral arms indicates that the arms have little effect on the star-forming productivity of dense clumps or on the average evolutionary state of star-forming clumps, and no change in the latter across the spiral arms where the line of sight is along a tangent. The latter might be surprising since a lag between dust/gas-traced and star-traced arms is predicted by the density-wave theory of spiral arms and has been reported several times in qualitative studies of nearby face-on galaxies, but not supported by more recent observational work \citep[see][and references therein]{Foyle2011}.

\begin{figure}
\includegraphics[width=0.5\textwidth]{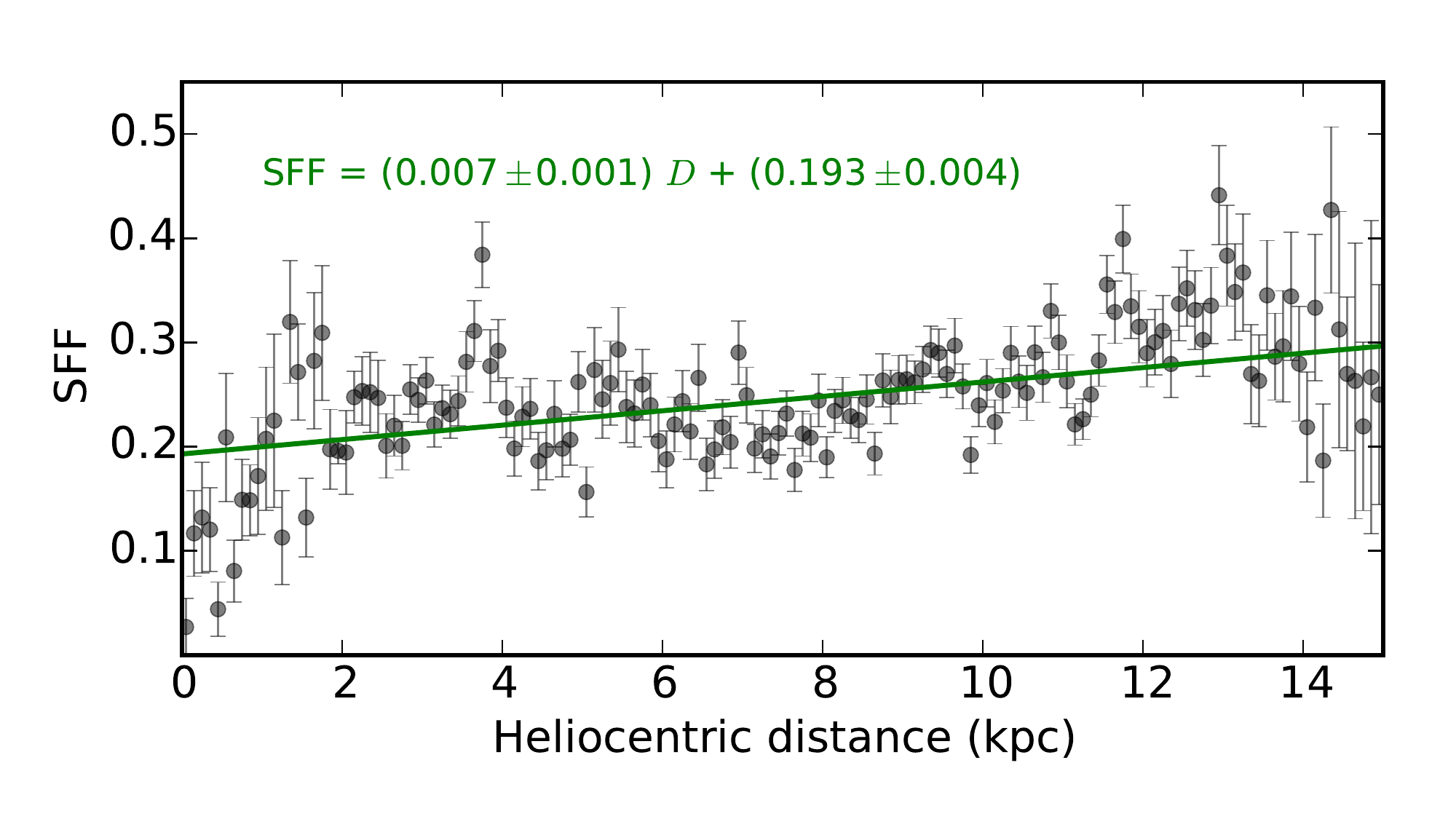}
\caption{\label{f:sff_heliodist} The star forming fraction (SFF) as a function of heliocentric distance in kiloparsecs. The result of a linear weighted least squares fit to the trend is shown in the green solid line, the slope of which is $+$0.007$\pm$0.001 per kpc.}
\end{figure}

\begin{figure}
\includegraphics[width=0.5\textwidth]{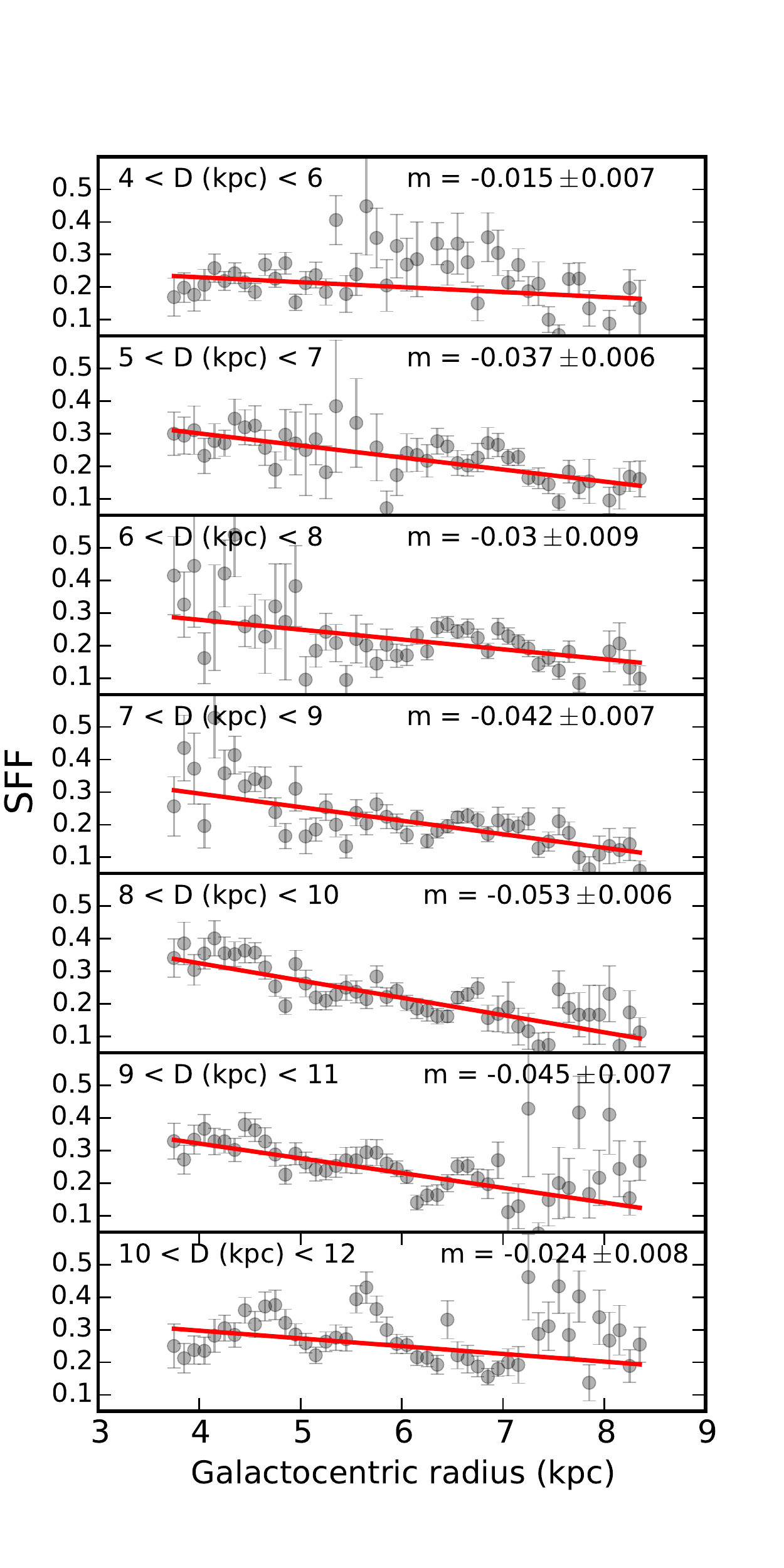}
\caption{\label{f:sff_heliodist_bias} SFF plotted as a function of $R_{\rm GC}$ using sources from 2\,kpc-wide heliocentric distance bins specified in the upper left of each panel. The best weighted least-squares linear fit is shown in the red line, the slope (m) of which is shown in each panel. Further statistics can be found in Table~\ref{t:sff_rgc_chunks}.}
\end{figure}

\begin{table}
\caption{SFF versus $R_{\rm GC}$ in variable heliocentric distance ranges. \label{t:sff_rgc_chunks}}
\begin{center}
\begin{tabular}{crcc}
$D$ range & $N_{\rm tot}$  & mean SFF & slope$^a$ \\
(kpc) & &  & (kpc$^{-1}$) \\
\hline
4  $< D <$  6 & 5865  & 0.226 & $-$0.015$\pm$0.007 \\
5  $< D <$  7 & 5832  & 0.227 & $-$0.037$\pm$0.006 \\
6  $< D <$  8 & 7937  & 0.214 & $-$0.030$\pm$0.009 \\
7  $< D <$  9 & 9802  & 0.217 & $-$0.042$\pm$0.007 \\
8  $< D <$ 10 & 10845 & 0.247 & $-$0.053$\pm$0.006 \\
9  $< D <$ 11 & 10529 & 0.264 & $-$0.045$\pm$0.007 \\
10 $< D <$ 12 & 9981  & 0.274 & $-$0.024$\pm$0.008 \\
\hline
\end{tabular}
\end{center}
$^a$ Slope of the SFF versus $R_{\rm GC}$ relation in 2\,kpc-wide heliocentric distance bins, as shown in Figure~\ref{f:sff_heliodist_bias}, fit between 3.1\,kpc $<$ $R_{\rm GC}$ $<$ 8.6\,kpc. 
\end{table}

\subsection{Potential biases in measuring the SFF}

As Hi-GAL provides us with an unprecedented number of uniformly-surveyed sources, we expect that any bias in our findings is distance-related.
Our study considers sources out to $D\sim$20\,kpc heliocentric distance, but given the longitude limits, this translates to a much smaller range of $R_{\rm GC}$, such that 96\% of these sources fall within the 3.1\,kpc $< R_{\rm GC} <$ 8.6\,kpc range used in the above analysis. The SFF as a function of heliocentric distance is shown in Figure~\ref{f:sff_heliodist}. There is a shallow but statistically significant slope of $+0.007 \pm 0.001$ which suggests a distance-related bias affecting the sample.

There are several possible effects at work here. First, the physical size corresponding to the resolution element increases with distance, such that (assuming a uniform distribution of sources on average) the number of sources overlapping with the beam will increase with distance and also the likelihood that one of those sources is 70\,$\mu$m-bright, tending to increase the SFF with heliocentric distance (see Figure \ref{f:sff_heliodist_bias} and Table \ref{t:sff_rgc_chunks}). Second, the typical spectral energy distributions of both starless and protostellar  Hi-GAL sources \citep[e.g.][]{Giannini2012} are intrinsically brighter at wavelengths longer than 160\micron. If a 70\micron counterpart is detected, it is typically a ``weaker'' (i.e. fewer $\sigma$ above rms) detection \citep[see Fig 3 in][]{Molinari2016b}. Thus, at large distances, sources are more readily detected at longer wavelengths, resulting in an increasing fraction of genuine protostellar sources being mis-classified as starless, effectively reducing the SFF with distance. 
Another potential related bias is the effect of distance on the average observed clump mass and luminosity of sources. At large distances, a higher fraction of the sources will be higher mass, which have shorter infrared-dark lifetimes \citep{Urquhart2014c}. In any case the gradient of this relationship is only one third of that in the relationship of SFF with $R_{\rm GC}$ and cannot be the cause of the latter, especially since the relationship between $R_{\rm GC}$ and $D$ is not a one-to-one correlation.

We can get a sense of the impact this shallow distance bias may have on the trend with $R_{\rm GC}$ by looking at the SFF versus $R_{\rm GC}$ using objects confined to narrow heliocentric distance bins. The SFF as a function of $R_{\rm GC}$ using sources within 2\,kpc intervals\footnote{The selection of the 2\,kpc width was made to ensure a good sample size ($N > 5000$) was available.} of heliocentric distance is shown in Figure~\ref{f:sff_heliodist_bias}. As expected from Figure~\ref{f:sff_heliodist}, the mean SFF increases slightly as the distance centre moves outward. We note that not only do all distance intervals show a significant anti-correlation, also the most populated distance intervals (covering distances between 7 and 11\,kpc) exhibit a steeper slope than the full sample value by a factor of $\sim$2, lending credibility to the overall robustness of the trend.

\begin{figure}
\includegraphics[width=0.5\textwidth]{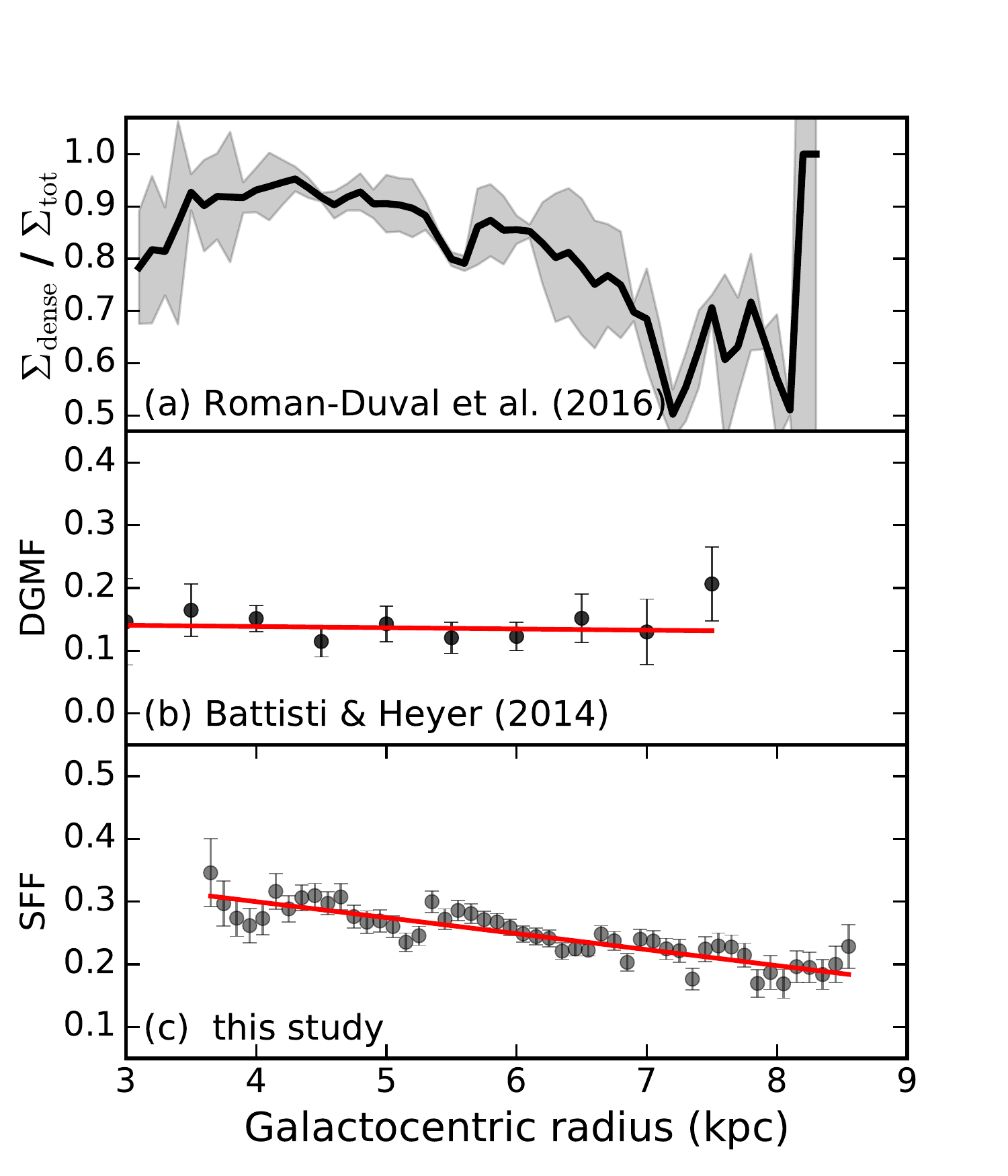}
\caption{{\bf (a)} The ratio of dense ($^{13}$CO-emitting) gas surface density to the total ($^{12}$CO + $^{13}$CO) gas surface density \citep[from Figure 12 of][]{Roman-Duval2016}. The black line shows the mean, and the light grey shaded area represents the total error budget. {\bf (b)} The median dense gas mass fraction (DGMF, defined as the ratio mass traced by sub-millimetre dust emission to the mass in the parent cloud traced by $^{13}$CO) reported in \citet{Battisti2014} as a function of $R_{\rm GC}$ plotted with the standard error of the DGMF in each 0.5\,kpc-wide bin. {\bf (c)} The SFF gradient with $R_{\rm GC}$ from Figure~\ref{f:Nsources_area_Rgc}. \label{f:radial_trends}}
\end{figure}

\subsection{What drives the gradient in SFF with $R_{\rm GC}$?}

It is far from clear what the physical origin of a gradual decline in SFF with $R_{\rm GC}$ over 5\,kpc might be. Since star formation is observed to be closely correlated with dense gas \citep[e.g.][]{Lada2010}, one might expect the SFF to be greater where the fraction of dense gas is higher. 
On kiloparsec scales, \citet{Roman-Duval2016} show that the fraction of ``dense'' gas -- defined as the fraction of mass in $^{13}$CO out of the ``total'' molecular mass (traced by $^{12}$CO + $^{13}$CO emission) -- does decline with $R_{\rm GC}$, roughly from 0.9 to 0.6 over the 3\,kpc $< R_{\rm GC} <$ 8\,kpc range (a gradient of $-$0.06\,kpc$^{-1}$, Figure~\ref{f:radial_trends}a). Within individual molecular clouds, however, the fraction of gas at even higher densities -- defined as the ratio of total mass in compact sub-millimetre clumps of dust emission to the total mass of the host cloud traced by $^{13}$CO -- shows no dependence on $R_{\rm GC}$ \citep[Figure~\ref{f:radial_trends}b, see also][]{Eden2013,Battisti2014}. This suggests that once molecular clouds form dense structures (which we observe as sub-millimetre or Hi-GAL clumps), the prevalence of star formation (or SFF) is governed by other internal properties, perhaps inherited from their environment. Below, we focus our discussion on the known large scale radial properties that have been observed in the Galactic disc including metallicity, radiation field, thermal and turbulent pressure and rotational shear.

The known negative metallicity gradient in the Galactic disc, when traced by HII regions and OB stars, is in the region of 0.06 -- 0.07 dex kpc$^{-1}$ within the approximate $R_{\rm GC}$ range covered in the present study \citep{Chiappini2001,Lepine2011b}, which translates to a reduction by a factor of 2 over 5\,kpc, while the measured SFF slope of $-$0.026 kpc$^{-1}$ produces only a 13\% decline in 5\,kpc. Reduced metallicity implies lower dust-to-gas ratio and reduced CO/H$_2$ abundance, and so less efficient cooling and turbulent energy dissipation. This might be expected to result in less efficient star formation. However, \citet{GloverClark2012c} predict that, while the fraction of total molecular cloud mass traced by CO may decrease, the star-formation rate within clouds has little sensitivity to the metallicity. \citet{Hocuk2016} also suggest that grain surface chemistry has only a small effect on star formation in molecular clouds. We cautiously note that the SFF traces star formation within clumps and not clouds and is independent of measured clump masses.

A radial decrease in radiation-field strength (in both photon intensity and hardness) should offset, to some extent, the reduced shielding that a declining metallicity produces via reduced dust and CO abundance \citep{Sandstrom2013}, so the destruction rate of both these will be less than expected from reduced metallicity alone \citep{GloverClark2012c}. Other potential effects of the radiation field related to star formation include changes in the ionisation fraction, a decrease in which may reduce magnetic-field support of clumps against gravity, and in thermal energy input to the ISM, but both effects are more likely to produce a positive SFF gradient than the observed negative one. 

\citet{Wolfire2003} estimate the typical thermal pressure in the Galactic plane interval 3\,kpc $<$ $R_{\rm GC}$ $< 18$\,kpc to be $P_{\rm therm}/k \simeq 1.4 \times 10^4\,\exp(-R/5.5\,\mbox{kpc})$\,K\,cm$^{-3}$ in the 3\,kpc $<$ $R_{\rm GC}$ $<$ 18\,kpc range  i.e. a shallow declining exponential. They predict a similar but flatter turbulent pressure gradient ($\propto \exp(-R/\mathrm{7.5\,{kpc}})$) between 3 and 10\,kpc. The SFF gradient with $R_{\rm GC}$ is much shallower, however the \citet{Wolfire2003} relation predicts a factor of 3.6 reduction in the pressure between 3\,kpc and 10\,kpc, corresponding to a linear gradient of 0.5$<P>$ per kpc. \citet{Rigby_PhD} find that, while the average thermal pressure in the denser parts of molecular clouds traced by $^{13}$CO ($J = 3\rightarrow 2$) is similar to that of the neutral gas, the turbulent pressures are higher by one or two orders of magnitude. On the other hand, the negative SFF gradient appears inconsistent with the proposition that increased turbulent pressure produces a raised density threshold for star formation \citep{Kruijssen2014}.

Rotational shear might be another suspect, contributing to turbulent pressure and the specific angular momentum of clouds and clumps, both of which may affect star-formation productivity. However, shear has been shown to have little effect on SFE within clouds \citep{Dib2012}. While shear is high at inner $R_{\rm GC}$, it decreases rapidly with increasing $R_{\rm GC}$ and is relatively flat and low beyond 3\,kpc, where the SFF decreases steadily. Again, the gradient appears to be in the wrong sense with high SFF where the shear is also higher.

As part of their investigation into the low star-formation efficiency in the CMZ, \citet{Kruijssen2014} suggest that the gravitational stability of the Galactic disc is increased inside $\sim$4\,kpc, due to the ratio of Toomre Q parameter to gas surface density. We might therefore expect the SFE (i.e. the conversion of total gas mass to stars) to decrease within this radius, but it is not clear how this might relate to the rate of production of stars in dense clumps measured by the SFF. \citet{Koda2016} show that the molecular gas fraction increases steadily with decreasing $R_{\rm GC}$, but we see in Figure~\ref{f:Nsources_area_Rgc} that the surface density of mass in dense clumps falls rapidly within 4\,kpc \citep{Bronfman1988,Urquhart2014a}. The production of molecular clouds from neutral gas therefore is more efficient at small $R_{\rm GC}$ where the H$_2$/HI ratio is nearly 100\% \citep{Koda2016}. The fraction of molecular gas in the form of dense clumps within these clouds, while more or less steady, on average outside $\sim$4\,kpc, albeit with very large, apparently random variations from cloud to cloud \citep{Eden2012,Eden2013}, falls sharply inside this radius.

Of the above mechanisms that might have a connection to the star-formation productivity of these dense clumps traced by Hi-GAL, most should affect the SFF in the opposite sense than is observed. Therefore the connection between the several-kpc-scale consistent gradient in SFF and environmental conditions is obscure, not least because the dense clumps, once formed, might be expected to go on to form stars independent of their environment.

\section{Summary}
We have examined Galactic scale trends in the distribution of Hi-GAL sources as a function of Galactocentric radius. We use the fraction of sources with a 70\micron counterpart (the so-called star-forming fraction, or SFF) as a measure of the prevalence of star formation in sources throughout the Galaxy. The mean SFF is 25\% in the range 3.1\,kpc $< R_{\rm GC} <$ 8.6\,kpc and decreases steadily as a function of $R_{\rm GC}$. A weighted least squares linear fit to the relation between SFF and $R_{\rm GC}$ yields a slope of $-$0.026$\pm$0.002 per kpc. Splitting the catalogue between the north and south yield similar (and robust) trends of $-$0.030$\pm$0.004 and $-$0.021$\pm$0.006 per kpc, respectively. Heliocentric distance-based biases can not account for the magnitude of the trend.

In the considered $R_{\rm GC}$ range, the spiral arms appear as features in the overall distribution of sources, but no convincing signal in SFF is evident at these $R_{\rm GC}$ locations. This is consistent with analogous efforts studying in trends in evolutionary stage (e.g. as probed by $L_{\rm IR}$/$M_{\rm clump}$) or clump formation efficiency, suggesting the SFF may be tied to these quantities. One notable difference in our study compared to these previous works is that we find no distinction is SFF at the Sagittarius spiral arm. In previous studies, it is seen as a peak in massive star formation signposts and gas temperature, but the SFF has no such peak. We speculate that this can be explained by the fact that the SFF does not account for local high-luminosity sources. Further study is needed on the source property distribution around the Sagittarius and all spiral arms in order to determine their role in Galactic scale star formation.

The SFF exhibits a negative gradient with $R_{\rm GC}$ despite the DGMF showing no such trend over the same $R_{\rm GC}$ range, indicating that the SFF may be weakly dependent on one or more large-scale environmental quantities, such as metallicity, radiation field, thermal pressure, turbulent pressure or shear, each of which exhibits some dependence on $R_{\rm GC}$. Considered individually, most of the Galactic-scale trends in these quantities would imply a positive gradient in SFF with $R_{\rm GC}$ rather than the observed negative trend. The interplay of these physical quantities across the Galaxy is clearly complex, and the SFF can serve as a useful benchmark for Galactic-scale simulations that test these phenomena. Moreover, further observational work, including an extension of this study to a larger range of $R_{\rm GC}$, will aid in understanding what drives this Galactic-scale trend. 

\section*{Acknowledgements}
The authors gratefully acknowledge Julia Roman-Duval for providing data for comparison and the anonymous referee for helpful suggestions. This research has made use of NASA Astrophysics Data System and Astropy, a community-developed core Python package for Astronomy \citep{astropy}. This work is part of the {\sc vialactea} Project, a Collaborative Project under Framework Programme 7 of the European Union, funded under Contract \# 607380.


\begin{thebibliography}{49}
\expandafter\ifx\csname natexlab\endcsname\relax\def\natexlab#1{#1}\fi

\bibitem[{{Astropy Collaboration} {et~al.}(2013){Astropy Collaboration},
  {Robitaille}, {Tollerud}, {Greenfield}, {Droettboom}, {Bray}, {Aldcroft},
  {Davis}, {Ginsburg}, {Price-Whelan}, {Kerzendorf}, {Conley}, {Crighton},
  {Barbary}, {Muna}, {Ferguson}, {Grollier}, {Parikh}, {Nair}, {Unther},
  {Deil}, {Woillez}, {Conseil}, {Kramer}, {Turner}, {Singer}, {Fox}, {Weaver},
  {Zabalza}, {Edwards}, {Azalee Bostroem}, {Burke}, {Casey}, {Crawford},
  {Dencheva}, {Ely}, {Jenness}, {Labrie}, {Lian Lim}, {Pierfederici},
  {Pontzen}, {Ptak}, {Refsdal}, {Servillat}, \& {Streicher}}]{astropy}
{Astropy Collaboration}, {Robitaille}, T.~P., {Tollerud}, E.~J., {et~al.} 2013,
  \aap, 558, A33

\bibitem[{{Battisti} \& {Heyer}(2014)}]{Battisti2014}
{Battisti}, A.~J. \& {Heyer}, M.~H. 2014, \apj, 780, 173

\bibitem[{{Bertoldi}(1989)}]{Bertoldi1989}
{Bertoldi}, F. 1989, \apj, 346, 735

\bibitem[{{Bisbas} {et~al.}(2009){Bisbas}, {W{\"u}nsch}, {Whitworth}, \&
  {Hubber}}]{Bisbas2009}
{Bisbas}, T.~G., {W{\"u}nsch}, R., {Whitworth}, A.~P., \& {Hubber}, D.~A. 2009,
  \aap, 497, 649

\bibitem[{{Bronfman} {et~al.}(1988){Bronfman}, {Cohen}, {Alvarez}, {May}, \&
  {Thaddeus}}]{Bronfman1988}
{Bronfman}, L., {Cohen}, R.~S., {Alvarez}, H., {May}, J., \& {Thaddeus}, P.
  1988, \apj, 324, 248

\bibitem[{{Chiappini} {et~al.}(2001){Chiappini}, {Matteucci}, \&
  {Romano}}]{Chiappini2001}
{Chiappini}, C., {Matteucci}, F., \& {Romano}, D. 2001, \apj, 554, 1044

\bibitem[{{Dame} {et~al.}(2001){Dame}, {Hartmann}, \& {Thaddeus}}]{Dame2001}
{Dame}, T.~M., {Hartmann}, D., \& {Thaddeus}, P. 2001, \apj, 547, 792

\bibitem[{{Dib} {et~al.}(2012){Dib}, {Helou}, {Moore}, {Urquhart}, \&
  {Dariush}}]{Dib2012}
{Dib}, S., {Helou}, G., {Moore}, T.~J.~T., {Urquhart}, J.~S., \& {Dariush}, A.
  2012, \apj, 758, 125

\bibitem[{{Eden} {et~al.}(2012){Eden}, {Moore}, {Plume}, \&
  {Morgan}}]{Eden2012}
{Eden}, D.~J., {Moore}, T.~J.~T., {Plume}, R., \& {Morgan}, L.~K. 2012, \mnras,
  422, 3178
  
\bibitem[{{Eden} {et~al.}(2013){Eden}, {Moore}, {Morgan}, {Thompson}, \&
  {Urquhart}}]{Eden2013}
{Eden}, D.~J., {Moore}, T.~J.~T., {Morgan}, L.~K., {Thompson}, M.~A., \&
  {Urquhart}, J.~S. 2013, \mnras, 431, 1587

\bibitem[{{Eden} {et~al.}(2015){Eden}, {Moore}, {Urquhart}, {Elia}, {Plume},
  {Rigby}, \& {Thompson}}]{Eden2015}
{Eden}, D.~J., {Moore}, T.~J.~T., {Urquhart}, J.~S., {et~al.} 2015, \mnras,
  452, 289

\bibitem[{{Foyle} {et~al.}(2011){Foyle}, {Rix}, {Dobbs}, {Leroy}, \&
  {Walter}}]{Foyle2011}
{Foyle}, K., {Rix}, H.-W., {Dobbs}, C.~L., {Leroy}, A.~K., \& {Walter}, F.
  2011, \apj, 735, 101

\bibitem[{{Giannini} {et~al.}(2012){Giannini}, {Elia}, {Lorenzetti},
  {Molinari}, {Motte}, {Schisano}, {Pezzuto}, {Pestalozzi}, {di Giorgio},
  {Andr{\'e}}, {Hill}, {Benedettini}, {Bontemps}, {di Francesco}, {Fallscheer},
  {Hennemann}, {Kirk}, {Minier}, {Nguyen Luong}, {Polychroni}, {Rygl},
  {Saraceno}, {Schneider}, {Spinoglio}, {Testi}, {Ward-Thompson}, \&
  {White}}]{Giannini2012}
{Giannini}, T., {Elia}, D., {Lorenzetti}, D., {et~al.} 2012, \aap, 539, A156

\bibitem[{{Glover} \& {Clark}(2012)}]{GloverClark2012c}
{Glover}, S.~C.~O. \& {Clark}, P.~C. 2012, \mnras, 426, 377

\bibitem[{{Griffin} {et~al.}(2010){Griffin}, {Abergel}, {Abreu}, {Ade},
  {Andr{\'e}}, {Augueres}, {Babbedge}, {Bae}, {Baillie}, {Baluteau}, {Barlow},
  {Bendo}, {Benielli}, {Bock}, {Bonhomme}, {Brisbin}, {Brockley-Blatt},
  {Caldwell}, {Cara}, {Castro-Rodriguez}, {Cerulli}, {Chanial}, {Chen},
  {Clark}, {Clements}, {Clerc}, {Coker}, {Communal}, {Conversi}, {Cox},
  {Crumb}, {Cunningham}, {Daly}, {Davis}, {de Antoni}, {Delderfield}, {Devin},
  {di Giorgio}, {Didschuns}, {Dohlen}, {Donati}, {Dowell}, {Dowell}, {Duband},
  {Dumaye}, {Emery}, {Ferlet}, {Ferrand}, {Fontignie}, {Fox}, {Franceschini},
  {Frerking}, {Fulton}, {Garcia}, {Gastaud}, {Gear}, {Glenn}, {Goizel},
  {Griffin}, {Grundy}, {Guest}, {Guillemet}, {Hargrave}, {Harwit}, {Hastings},
  {Hatziminaoglou}, {Herman}, {Hinde}, {Hristov}, {Huang}, {Imhof}, {Isaak},
  {Israelsson}, {Ivison}, {Jennings}, {Kiernan}, {King}, {Lange}, {Latter},
  {Laurent}, {Laurent}, {Leeks}, {Lellouch}, {Levenson}, {Li}, {Li},
  {Lilienthal}, {Lim}, {Liu}, {Lu}, {Madden}, {Mainetti}, {Marliani}, {McKay},
  {Mercier}, {Molinari}, {Morris}, {Moseley}, {Mulder}, {Mur}, {Naylor},
  {Nguyen}, {O'Halloran}, {Oliver}, {Olofsson}, {Olofsson}, {Orfei}, {Page},
  {Pain}, {Panuzzo}, {Papageorgiou}, {Parks}, {Parr-Burman}, {Pearce},
  {Pearson}, {P{\'e}rez-Fournon}, {Pinsard}, {Pisano}, {Podosek}, {Pohlen},
  {Polehampton}, {Pouliquen}, {Rigopoulou}, {Rizzo}, {Roseboom}, {Roussel},
  {Rowan-Robinson}, {Rownd}, {Saraceno}, {Sauvage}, {Savage}, {Savini},
  {Sawyer}, {Scharmberg}, {Schmitt}, {Schneider}, {Schulz}, {Schwartz},
  {Shafer}, {Shupe}, {Sibthorpe}, {Sidher}, {Smith}, {Smith}, {Smith},
  {Spencer}, {Stobie}, {Sudiwala}, {Sukhatme}, {Surace}, {Stevens}, {Swinyard},
  {Trichas}, {Tourette}, {Triou}, {Tseng}, {Tucker}, {Turner}, {Vaccari},
  {Valtchanov}, {Vigroux}, {Virique}, {Voellmer}, {Walker}, {Ward}, {Waskett},
  {Weilert}, {Wesson}, {White}, {Whitehouse}, {Wilson}, {Winter}, {Woodcraft},
  {Wright}, {Xu}, {Zavagno}, {Zemcov}, {Zhang}, \&
  {Zonca}}]{A&ASpecialIssue-SPIRE}
{Griffin}, M.~J., {Abergel}, A., {Abreu}, A., {et~al.} 2010, \aap, 518, L3

\bibitem[{{Heiles} \& {Troland}(2005)}]{HeilesTroland2005}
{Heiles}, C. \& {Troland}, T.~H. 2005, \apj, 624, 773

\bibitem[{{Heyer} {et~al.}(1998){Heyer}, {Brunt}, {Snell}, {Howe}, {Schloerb},
  \& {Carpenter}}]{Heyer1998}
{Heyer}, M.~H., {Brunt}, C., {Snell}, R.~L., {et~al.} 1998, \apjs, 115, 241

\bibitem[{{Hocuk} {et~al.}(2016){Hocuk}, {Cazaux}, {Spaans}, \&
  {Caselli}}]{Hocuk2016}
{Hocuk}, S., {Cazaux}, S., {Spaans}, M., \& {Caselli}, P. 2016, \mnras, 456,
  2586

\bibitem[{{Hou} \& {Han}(2014)}]{HouHan2014}
{Hou}, L.~G. \& {Han}, J.~L. 2014, \aap, 569, A125


\bibitem[Koda et al.(2016)]{Koda2016} Koda, J., Scoville, N.,
 \& Heyer, M.\ 2016, \apj, 823, 76 


\bibitem[{{Kruijssen} {et~al.}(2014){Kruijssen}, {Longmore}, {Elmegreen},
  {Murray}, {Bally}, {Testi}, \& {Kennicutt}}]{Kruijssen2014}
{Kruijssen}, J.~M.~D., {Longmore}, S.~N., {Elmegreen}, B.~G., {et~al.} 2014,
  \mnras, 440, 3370

\bibitem[Lada et al.(2010)]{Lada2010} Lada, C.~J., Lombardi, M., \& Alves, J.~F.\ 2010, \apj, 724, 687 


\bibitem[{{L{\'e}pine} {et~al.}(2011){L{\'e}pine}, {Cruz}, {Scarano}, {Barros},
  {Dias}, {Pomp{\'e}ia}, {Andrievsky}, {Carraro}, \& {Famaey}}]{Lepine2011b}
{L{\'e}pine}, J.~R.~D., {Cruz}, P., {Scarano}, Jr., S., {et~al.} 2011, \mnras,
  417, 698

\bibitem[{{Molinari} {et~al.}(2008){Molinari}, {Pezzuto}, {Cesaroni}, {Brand},
  {Faustini}, \& {Testi}}]{Molinari2008}
{Molinari}, S., {Pezzuto}, S., {Cesaroni}, R., {et~al.} 2008, \aap, 481, 345

\bibitem[{{Molinari} {et~al.}(2010{\natexlab{a}}){Molinari}, {Swinyard},
  {Bally}, {Barlow}, {Bernard}, {Martin}, {Moore}, {Noriega-Crespo}, {Plume},
  {Testi}, {Zavagno}, {Abergel}, {Ali}, {Anderson}, {Andr{\'e}}, {Baluteau},
  {Battersby}, {Beltr{\'a}n}, {Benedettini}, {Billot}, {Blommaert}, {Bontemps},
  {Boulanger}, {Brand}, {Brunt}, {Burton}, {Calzoletti}, {Carey}, {Caselli},
  {Cesaroni}, {Cernicharo}, {Chakrabarti}, {Chrysostomou}, {Cohen},
  {Compiegne}, {de Bernardis}, {de Gasperis}, {di Giorgio}, {Elia}, {Faustini},
  {Flagey}, {Fukui}, {Fuller}, {Ganga}, {Garcia-Lario}, {Glenn}, {Goldsmith},
  {Griffin}, {Hoare}, {Huang}, {Ikhenaode}, {Joblin}, {Joncas}, {Juvela},
  {Kirk}, {Lagache}, {Li}, {Lim}, {Lord}, {Marengo}, {Marshall}, {Masi},
  {Massi}, {Matsuura}, {Minier}, {Miville-Desch{\^e}nes}, {Montier}, {Morgan},
  {Motte}, {Mottram}, {M{\"u}ller}, {Natoli}, {Neves}, {Olmi}, {Paladini},
  {Paradis}, {Parsons}, {Peretto}, {Pestalozzi}, {Pezzuto}, {Piacentini},
  {Piazzo}, {Polychroni}, {Pomar{\`e}s}, {Popescu}, {Reach}, {Ristorcelli},
  {Robitaille}, {Robitaille}, {Rod{\'o}n}, {Roy}, {Royer}, {Russeil},
  {Saraceno}, {Sauvage}, {Schilke}, {Schisano}, {Schneider}, {Schuller},
  {Schulz}, {Sibthorpe}, {Smith}, {Smith}, {Spinoglio}, {Stamatellos},
  {Strafella}, {Stringfellow}, {Sturm}, {Taylor}, {Thompson}, {Traficante},
  {Tuffs}, {Umana}, {Valenziano}, {Vavrek}, {Veneziani}, {Viti}, {Waelkens},
  {Ward-Thompson}, {White}, {Wilcock}, {Wyrowski}, {Yorke}, \&
  {Zhang}}]{Molinari2010b}
{Molinari}, S., {Swinyard}, B., {Bally}, J., {et~al.} 2010{\natexlab{a}}, \aap,
  518, L100

\bibitem[{{Molinari} {et~al.}(2010{\natexlab{b}}){Molinari}, {Swinyard},
  {Bally}, {Barlow}, {Bernard}, {Martin}, {Moore}, {Noriega-Crespo}, {Plume},
  {Testi}, {Zavagno}, {Abergel}, {Ali}, {Andr{\'e}}, {Baluteau}, {Benedettini},
  {Bern{\'e}}, {Billot}, {Blommaert}, {Bontemps}, {Boulanger}, {Brand},
  {Brunt}, {Burton}, {Campeggio}, {Carey}, {Caselli}, {Cesaroni}, {Cernicharo},
  {Chakrabarti}, {Chrysostomou}, {Codella}, {Cohen}, {Compiegne}, {Davis}, {de
  Bernardis}, {de Gasperis}, {Di Francesco}, {di Giorgio}, {Elia}, {Faustini},
  {Fischera}, {Fukui}, {Fuller}, {Ganga}, {Garcia-Lario}, {Giard}, {Giardino},
  {Glenn}, {Goldsmith}, {Griffin}, {Hoare}, {Huang}, {Jiang}, {Joblin},
  {Joncas}, {Juvela}, {Kirk}, {Lagache}, {Li}, {Lim}, {Lord}, {Lucas},
  {Maiolo}, {Marengo}, {Marshall}, {Masi}, {Massi}, {Matsuura}, {Meny},
  {Minier}, {Miville-Desch{\^e}nes}, {Montier}, {Motte}, {M{\"u}ller},
  {Natoli}, {Neves}, {Olmi}, {Paladini}, {Paradis}, {Pestalozzi}, {Pezzuto},
  {Piacentini}, {Pomar{\`e}s}, {Popescu}, {Reach}, {Richer}, {Ristorcelli},
  {Roy}, {Royer}, {Russeil}, {Saraceno}, {Sauvage}, {Schilke},
  {Schneider-Bontemps}, {Schuller}, {Schultz}, {Shepherd}, {Sibthorpe},
  {Smith}, {Smith}, {Spinoglio}, {Stamatellos}, {Strafella}, {Stringfellow},
  {Sturm}, {Taylor}, {Thompson}, {Tuffs}, {Umana}, {Valenziano}, {Vavrek},
  {Viti}, {Waelkens}, {Ward-Thompson}, {White}, {Wyrowski}, {Yorke}, \&
  {Zhang}}]{Molinari2010a}
{Molinari}, S., {Swinyard}, B., {Bally}, J., {et~al.} 2010{\natexlab{b}},
  \pasp, 122, 314

\bibitem[Molinari et al.(2016)]{Molinari2016b} Molinari, S., Schisano, E., 
Elia, D., et al.\ 2016, \aap, 591, A149 

\bibitem[{{Moore} {et~al.}(2012){Moore}, {Urquhart}, {Morgan}, \&
  {Thompson}}]{Moore2012}
{Moore}, T.~J.~T., {Urquhart}, J.~S., {Morgan}, L.~K., \& {Thompson}, M.~A.
  2012, \mnras, 426, 701
  
\bibitem[{{Moore} {et~al.}(2015){Moore}, {Plume}, {Thompson}, {Parsons},
  {Urquhart}, {Eden}, {Dempsey}, {Morgan}, {Thomas}, {Buckle}, {Brunt},
  {Butner}, {Carretero}, {Chrysostomou}, {deVilliers}, {Fich}, {Hoare},
  {Manser}, {Mottram}, {Natario}, {Olguin}, {Peretto}, {Polychroni}, {Redman},
  {Rigby}, {Salji}, {Summers}, {Berry}, {Currie}, {Jenness}, {Pestalozzi},
  {Traficante}, {Bastien}, {diFrancesco}, {Davis}, {Evans}, {Friberg},
  {Fuller}, {Gibb}, {Gibson}, {Hill}, {Johnstone}, {Joncas}, {Longmore},
  {Lumsden}, {Martin}, {Lu'o'ng}, {Pineda}, {Purcell}, {Richer}, {Schieven},
  {Shipman}, {Spaans}, {Taylor}, {Viti}, {Weferling}, {White}, \&
  {Zhu}}]{Moore2015}
{Moore}, T.~J.~T., {Plume}, R., {Thompson}, M.~A., {et~al.} 2015, \mnras, 453,
  4264

\bibitem[{{Motte} {et~al.}(2007){Motte}, {Bontemps}, {Schilke}, {Schneider},
  {Menten}, \& {Brogui{\`e}re}}]{Motte_cygX}
{Motte}, F., {Bontemps}, S., {Schilke}, P., {et~al.} 2007, \aap, 476, 1243

\bibitem[{{Poglitsch} {et~al.}(2010){Poglitsch}, {Waelkens}, {Geis},
  {Feuchtgruber}, {Vandenbussche}, {Rodriguez}, {Krause}, {Renotte}, {van
  Hoof}, {Saraceno}, {Cepa}, {Kerschbaum}, {Agn{\`e}se}, {Ali}, {Altieri},
  {Andreani}, {Augueres}, {Balog}, {Barl}, {Bauer}, {Belbachir}, {Benedettini},
  {Billot}, {Boulade}, {Bischof}, {Blommaert}, {Callut}, {Cara}, {Cerulli},
  {Cesarsky}, {Contursi}, {Creten}, {De Meester}, {Doublier}, {Doumayrou},
  {Duband}, {Exter}, {Genzel}, {Gillis}, {Gr{\"o}zinger}, {Henning},
  {Herreros}, {Huygen}, {Inguscio}, {Jakob}, {Jamar}, {Jean}, {de Jong},
  {Katterloher}, {Kiss}, {Klaas}, {Lemke}, {Lutz}, {Madden}, {Marquet},
  {Martignac}, {Mazy}, {Merken}, {Montfort}, {Morbidelli}, {M{\"u}ller},
  {Nielbock}, {Okumura}, {Orfei}, {Ottensamer}, {Pezzuto}, {Popesso},
  {Putzeys}, {Regibo}, {Reveret}, {Royer}, {Sauvage}, {Schreiber}, {Stegmaier},
  {Schmitt}, {Schubert}, {Sturm}, {Thiel}, {Tofani}, {Vavrek}, {Wetzstein},
  {Wieprecht}, \& {Wiezorrek}}]{A&ASpecialIssue-PACS}
{Poglitsch}, A., {Waelkens}, C., {Geis}, N., {et~al.} 2010, \aap, 518, L2

\bibitem[{{Ragan} {et~al.}(2012){Ragan}, {Henning}, {Krause}, {Pitann},
  {Beuther}, {Linz}, {Tackenberg}, {Balog}, {Hennemann}, {Launhardt}, {Lippok},
  {Nielbock}, {Schmiedeke}, {Schuller}, {Steinacker}, {Stutz}, \&
  {Vasyunina}}]{Ragan2012b}
{Ragan}, S.~E., {Henning}, T., {Krause}, O., {et~al.} 2012, \aap, 547, A49

\bibitem[{{Ragan} {et~al.}(2014){Ragan}, {Henning}, {Tackenberg}, {Beuther},
  {Johnston}, {Kainulainen}, \& {Linz}}]{Ragan2014}
{Ragan}, S.~E., {Henning}, T., {Tackenberg}, J., {et~al.} 2014, \aap, 568, A73

\bibitem[{{Reid} {et~al.}(2014){Reid}, {Menten}, {Brunthaler}, {Zheng}, {Dame},
  {Xu}, {Wu}, {Zhang}, {Sanna}, {Sato}, {Hachisuka}, {Choi}, {Immer},
  {Moscadelli}, {Rygl}, \& {Bartkiewicz}}]{Reid2014a}
{Reid}, M.~J., {Menten}, K.~M., {Brunthaler}, A., {et~al.} 2014, \apj, 783, 130

\bibitem[Reid et al.(2016)]{Reid2016} Reid, M.~J., Dame, T.~M., 
Menten, K.~M., \& Brunthaler, A.\ 2016, \apj, 823, 77 

\bibitem[{{Rigby} {et~al.}(2016){Rigby}, {Moore}, {Plume}, {Eden}, {Urquhart},
  {Thompson}, {Mottram}, {Brunt}, {Butner}, {Dempsey}, {Gibson}, {Hatchell},
  {Jenness}, {Kuno}, {Longmore}, {Morgan}, {Polychroni}, {Thomas}, {White}, \&
  {Zhu}}]{Rigby2016}
{Rigby}, A.~J., {Moore}, T.~J.~T., {Plume}, R., {et~al.} 2016, \mnras, 456,
  2885

\bibitem[Rigby (2016)]{Rigby_PhD} Rigby, A.~J. 2016, Ph.D.~Thesis

\bibitem[{{Robitaille} {et~al.}(2012){Robitaille}, {Churchwell}, {Benjamin},
  {Whitney}, {Wood}, {Babler}, \& {Meade}}]{Robitaille2012}
{Robitaille}, T.~P., {Churchwell}, E., {Benjamin}, R.~A., {et~al.} 2012, \aap,
  545, A39

\bibitem[{{Roman-Duval} {et~al.}(2010){Roman-Duval}, {Jackson}, {Heyer},
  {Rathborne}, \& {Simon}}]{Roman-Duval2010}
{Roman-Duval}, J., {Jackson}, J.~M., {Heyer}, M., {Rathborne}, J., \& {Simon},
  R. 2010, \apj, 723, 492

\bibitem[Roman-Duval et al.(2016)]{Roman-Duval2016} Roman-Duval, J., 
 Heyer, M., Brunt, C.~M., et al.\ 2016, \apj, 818, 144 

\bibitem[{{Russeil} {et~al.}(2011){Russeil}, {Pestalozzi}, {Mottram},
  {Bontemps}, {Anderson}, {Zavagno}, {Beltr{\'a}n}, {Bally}, {Brand}, {Brunt},
  {Cesaroni}, {Joncas}, {Marshall}, {Martin}, {Massi}, {Molinari}, {Moore},
  {Noriega-Crespo}, {Olmi}, {Thompson}, {Wienen}, \& {Wyrowski}}]{Russeil2011}
{Russeil}, D., {Pestalozzi}, M., {Mottram}, J.~C., {et~al.} 2011, \aap, 526,
  A151

\bibitem[{{Sandstrom} {et~al.}(2013){Sandstrom}, {Leroy}, {Walter}, {Bolatto},
  {Croxall}, {Draine}, {Wilson}, {Wolfire}, {Calzetti}, {Kennicutt}, {Aniano},
  {Donovan Meyer}, {Usero}, {Bigiel}, {Brinks}, {de Blok}, {Crocker}, {Dale},
  {Engelbracht}, {Galametz}, {Groves}, {Hunt}, {Koda}, {Kreckel}, {Linz},
  {Meidt}, {Pellegrini}, {Rix}, {Roussel}, {Schinnerer}, {Schruba}, {Schuster},
  {Skibba}, {van der Laan}, {Appleton}, {Armus}, {Brandl}, {Gordon}, {Hinz},
  {Krause}, {Montiel}, {Sauvage}, {Schmiedeke}, {Smith}, \&
  {Vigroux}}]{Sandstrom2013}
{Sandstrom}, K.~M., {Leroy}, A.~K., {Walter}, F., {et~al.} 2013, \apj, 777, 5

\bibitem[{{Schuller} {et~al.}(2009){Schuller}, {Menten}, {Contreras},
  {Wyrowski}, {Schilke}, {Bronfman}, {Henning}, {Walmsley}, {Beuther},
  {Bontemps}, {Cesaroni}, {Deharveng}, {Garay}, {Herpin}, {Lefloch}, {Linz},
  {Mardones}, {Minier}, {Molinari}, {Motte}, {Nyman}, {Reveret}, {Risacher},
  {Russeil}, {Schneider}, {Testi}, {Troost}, {Vasyunina}, {Wienen}, {Zavagno},
  {Kovacs}, {Kreysa}, {Siringo}, \& {Wei{\ss}}}]{ATLASGAL}
{Schuller}, F., {Menten}, K.~M., {Contreras}, Y., {et~al.} 2009, \aap, 504, 415

\bibitem[{{Sodroski} {et~al.}(1997){Sodroski}, {Odegard}, {Arendt}, {Dwek},
  {Weiland}, {Hauser}, \& {Kelsall}}]{Sodroski1997}
{Sodroski}, T.~J., {Odegard}, N., {Arendt}, R.~G., {et~al.} 1997, \apj, 480,
  173

\bibitem[{{Traficante} {et~al.}(2015){Traficante}, {Fuller}, {Peretto},
  {Pineda}, \& {Molinari}}]{Traficante2015}
{Traficante}, A., {Fuller}, G.~A., {Peretto}, N., {Pineda}, J.~E., \&
  {Molinari}, S. 2015, \mnras, 451, 3089

\bibitem[{{Urquhart} {et~al.}(2013){Urquhart}, {Moore}, {Schuller}, {Wyrowski},
  {Menten}, {Thompson}, {Csengeri}, {Walmsley}, {Bronfman}, \&
  {K{\"o}nig}}]{Urquhart2013a}
{Urquhart}, J.~S., {Moore}, T.~J.~T., {Schuller}, F., {et~al.} 2013, \mnras,
  431, 1752
  
\bibitem[{{Urquhart} {et~al.}(2014{\natexlab{a}}){Urquhart}, {Figura}, {Moore},
  {Hoare}, {Lumsden}, {Mottram}, {Thompson}, \& {Oudmaijer}}]{Urquhart2014a}
{Urquhart}, J.~S., {Figura}, C.~C., {Moore}, T.~J.~T., {et~al.}
  2014{\natexlab{a}}, \mnras, 437, 1791

\bibitem[{{Urquhart} {et~al.}(2014{\natexlab{b}}){Urquhart}, {Moore},
  {Csengeri}, {Wyrowski}, {Schuller}, {Hoare}, {Lumsden}, {Mottram},
  {Thompson}, {Menten}, {Walmsley}, {Bronfman}, {Pfalzner}, {K{\"o}nig}, \&
  {Wienen}}]{Urquhart2014c}
{Urquhart}, J.~S., {Moore}, T.~J.~T., {Csengeri}, T., {et~al.}
  2014{\natexlab{b}}, \mnras, 443, 1555

\bibitem[{{Vall{\'e}e}(2014{\natexlab{a}})}]{Vallee2014c}
{Vall{\'e}e}, J.~P. 2014{\natexlab{a}}, \apjs, 215, 1

\bibitem[{{Vall{\'e}e}(2014{\natexlab{b}})}]{Vallee2014a}
{Vall{\'e}e}, J.~P. 2014{\natexlab{b}}, \aj, 148, 5

\bibitem[{{Wolfire} {et~al.}(2003){Wolfire}, {McKee}, {Hollenbach}, \&
  {Tielens}}]{Wolfire2003}
{Wolfire}, M.~G., {McKee}, C.~F., {Hollenbach}, D., \& {Tielens}, A.~G.~G.~M.
  2003, \apj, 587, 278

\end{thebibliography}
\end{document}